\documentclass{nature_with_images}

\newcommand{\be}{\begin{equation}}
\newcommand{\ee}{\end{equation}}
\newcommand{\bea}{\begin{eqnarray}}
\newcommand{\eea}{\end{eqnarray}}

\newcommand{\bJ}{{\bf J}}
\newcommand{\bj}{{\bf j}}
\newcommand{\bE}{{\bf E}}
\newcommand{\bx}{{\bf x}}

\newcommand{\ket}[1]{\left|#1\right>}
\newcommand{\bra}[1]{\left<#1\right|}

\newcommand{\expect}[1]{\left<#1\right>}

\newcommand{\cE}{{\cal E}}

\newcommand{\cP}{{\cal P}}

\newcommand{\var}{{\rm var}}

\usepackage{amsmath,bbm,amssymb}
\usepackage{siunitx,verbatim}
\usepackage{graphicx,color}

\title{Interaction-based quantum metrology showing scaling
beyond the Heisenberg limit.}

\author{M. Napolitano$^{1}$, M. Koschorreck$^{1}$,  B. Dubost$^{1,2}$,
N. Behbood$^1$, R. J. Sewell$^1$ \& M. W. Mitchell$^1$}

\newcommand{\OpS}{{\hat{S}}}
\newcommand{\Ops}{{\hat{s}}}
\newcommand{\OpH}{{\hat{H}}}
\newcommand{\OpF}{{\hat{F}}}
\newcommand{\NumPhot}{N_{\rm NL}}
\newcommand{\NumGen}{N}
\newcommand{\corrplot}{{$\philin,\phinl$ correlation plot}}
\newcommand{\corrplots}{{$\philin,\phinl$ correlation plots}}

\newcommand{\UpSym}{{\uparrow}}
\newcommand{\DownSym}{{\downarrow}}

\newcommand{\myexpect}[1]{\left<\right.#1\left.\right>}

\renewcommand{\cE}{{\vec{\cal E}}}
\renewcommand{\cP}{{\vec{\cal P}}}

\newcommand{\rb}{\ensuremath{^{87}\mathrm{Rb}}}
\newcommand{\phinl}{\ensuremath{\phi_{\rm{NL}}}}

\newcommand{\philin}{\ensuremath{\phi_{\rm{L}}}}

\newcommand{\dphiexpt}{\ensuremath{\Delta{\phi}_{\rm{NL}}}}

\newcommand{\unknown}{{\cal X}}
\newcommand{\introunk}{{\unknown}}
\newcommand{\rotstr}{{\unknown}}

\newcommand{\PerRtHz}{{\rm Hz}^{-1/2}}

\begin{document}

\maketitle

\begin{affiliations}
 \item ICFO-Institut de Ciencies Fotoniques, Mediterranean Technology Park,  
08860 Castelldefels (Barcelona), Spain. 
 \item Laboratoire Mat\'{e}riaux et Ph\'{e}nom\`{e}nes Quantiques, Universit\'{e} Paris Diderot et CNRS, \\UMR 7162, B\^{a}t. Condorcet, 75205 Paris Cedex 13, France.
\end{affiliations}

\begin{abstract}

Quantum metrology studies the use of entanglement and other
quantum resources to improve precision
measurement\cite{GiovannettiPRL2006}.  An interferometer using $N$
independent particles to measure a parameter $\introunk$ can
achieve at best the ``standard quantum limit'' (SQL) of
sensitivity $\delta \introunk \propto {N}^{-1/2}$.  The same
interferometer\cite{LeeJMO2002} using $N$ entangled particles can
achieve in principle the ``Heisenberg limit'' $\delta \introunk
\propto {N}^{-1}$, using exotic states\cite{MitchellN2004}. Recent
theoretical work 
argues that {\it
interactions} among particles may be a valuable resource for
quantum metrology, allowing scaling beyond the Heisenberg
limit\cite{BoixoPRL2007,ChoiPRA2008, RoyPRL2008}. Specifically, a
$k$-particle interaction will produce sensitivity $\delta
\introunk \propto N^{-k}$ with appropriate entangled states and
$\delta \introunk \propto N^{-(k-1/2)}$ even without
entanglement\cite{BoixoPRL2008}.  Here we demonstrate this
``super-Heisenberg'' scaling in a nonlinear,
non-destructive\cite{KoschorreckPRL2010a,KoschorreckPRL2010b}
measurement of the magnetisation\cite{KominisN2003,BudkerNP2007}
of an atomic ensemble\cite{HammererRMP2010}. We use fast optical
nonlinearities to generate a pairwise photon-photon
interaction\cite{NapolitanoJP2010} ($k=2$) while preserving
quantum-noise-limited
performance\cite{FleischhauerPRA2000,BoixoPRL2008}, to produce
$\delta \introunk \propto \NumGen^{-3/2}$. We observe
super-Heisenberg scaling over two orders of magnitude in $N$,
limited at large $N$ by higher-order nonlinear effects, in good
agreement with theory\cite{NapolitanoJP2010}.
For a measurement of limited duration, super-Heisenberg scaling allows the nonlinear
measurement to overtake in sensitivity a comparable linear
measurement with the same number of photons. In other scenarios,
however, higher-order nonlinearities prevent this crossover from
occurring, reflecting the subtle relationship of scaling to
sensitivity in nonlinear systems. This work shows that
inter-particle interactions can
improve sensitivity in a quantum-limited measurement, and
introduces a fundamentally new resource for quantum metrology.


\end{abstract}

The best instruments are interferometric in nature, and operate
according to the laws of quantum mechanics.  A collection of
particles, e.g.{,} photons or atoms, is prepared in a
superposition state, allowed to evolve under the action of a
Hamiltonian containing an unknown parameter $\unknown$, and
measured in agreement with quantum measurement theory. The
complementarity of quantum measurements\cite{ScullyNAT1991}
determines the ultimate sensitivity of these instruments.

Here we describe polarisation interferometry, used for example in
optical magnetometry to detect atomic
magnetisation\cite{BudkerNP2007 , WasilewskiPRL2010
,WolfgrammPRL2010}; similar theory describes other
interferometers\cite{LeeJMO2002}. A collection of $\NumGen$
photons, with circular plus/minus polarisations $\ket{+},\ket{-}$
is described by single-photon Stokes operators
$\Ops_i = \frac{1}{2}(\ket{+},\ket{-}) \sigma_i
(\bra{+},\bra{-})^T$, where the $\sigma_i$ are the Pauli matrices
and $\sigma_0$ is the identity. In traditional quantum metrology,
a Hamiltonian of the form $\hat{H} = \hbar \rotstr
\sum_{j=1}^{\NumGen} \Ops_z^{(j)}$ uniformly and independently
couples the photons to $\rotstr$, the parameter to be
measured\cite{GiovannettiPRL2006}. If the input state consists of
independent photons, the possible precision scales as $\delta
\rotstr \propto \NumGen^{-1/2}$, the shot-noise or standard
quantum limit (SQL). The $\NumGen^{-1/2}$ factor reflects the
statistical averaging of independent results. In contrast,
entangled states can be highly, even perfectly, correlated, giving
precision limited by $\delta \rotstr \propto \NumGen^{-1}$, the
Heisenberg limit (HL).

The above Hamiltonian is conveniently written $\hat{H} =
\hbar \rotstr \OpS_z$, where $\OpS_i \equiv \sum_{j=1}^{\NumGen}
\Ops_i^{(j)}$ is a collective variable describing the net
polarisation of the photons. The independence of the photons
manifests itself in the linearity of this Hamiltonian.  Recently
Boixo {\it et al.} have shown that interactions among particles,
or equivalently nonlinear Hamiltonians, can contribute to
measurement sensitivity and {give scaling} beyond the Heisenberg
limit\cite{BoixoPRL2007}. For example, a Hamiltonian $\OpH = \hbar
\rotstr \OpS_z^k$, i.e., with a $k$-order nonlinearity in
$\hat{\bf S}$, contains $k$-photon interaction terms
$\Ops_z^{(j_1)} \otimes \Ops_z^{(j_2)} \otimes \ldots \otimes
\Ops_z^{(j_k)}$. The number of such terms, and thus the signal
strength, grows as $\NumGen^k$, while the quantum noise
from the input states is unchanged. As a result, a
sensitivity limit of $\delta \rotstr \propto \NumGen^{-k}$ applies
when entanglement is used, and $\delta \rotstr \propto
\NumGen^{-(k-1/2)}$ in the absence of
entanglement\cite{BoixoPRL2008}.  For $k\ge 2$, this already gives a scaling better than the
Heisenberg limit, so-called ``super-Heisenberg'' (SH)
scaling\cite{BoixoPRL2008}. Note that interactions and
entanglement are compatible and both improve the scaling.  The
predicted advantage applies generally to quantum interferometry,
and proposed mechanisms to produce metrologically-relevant
interactions include Kerr nonlinearities\cite{BeltranPRA2005},
 {cold collisions in condensed atomic gases}\cite{BoixoPRL2008}, Duffing
nonlinearity in nano-mechanical resonators\cite{WoolleyNJP2008}
and a two-pass effective nonlinearity with an atomic
ensemble\cite{ChasePRA2009}.  Topological excitations in
nonlinear systems may also give advantageous scaling\cite{NegrettiPRA2008}.





\begin{figure}[h]
    \centering
    \includegraphics[width=\textwidth]{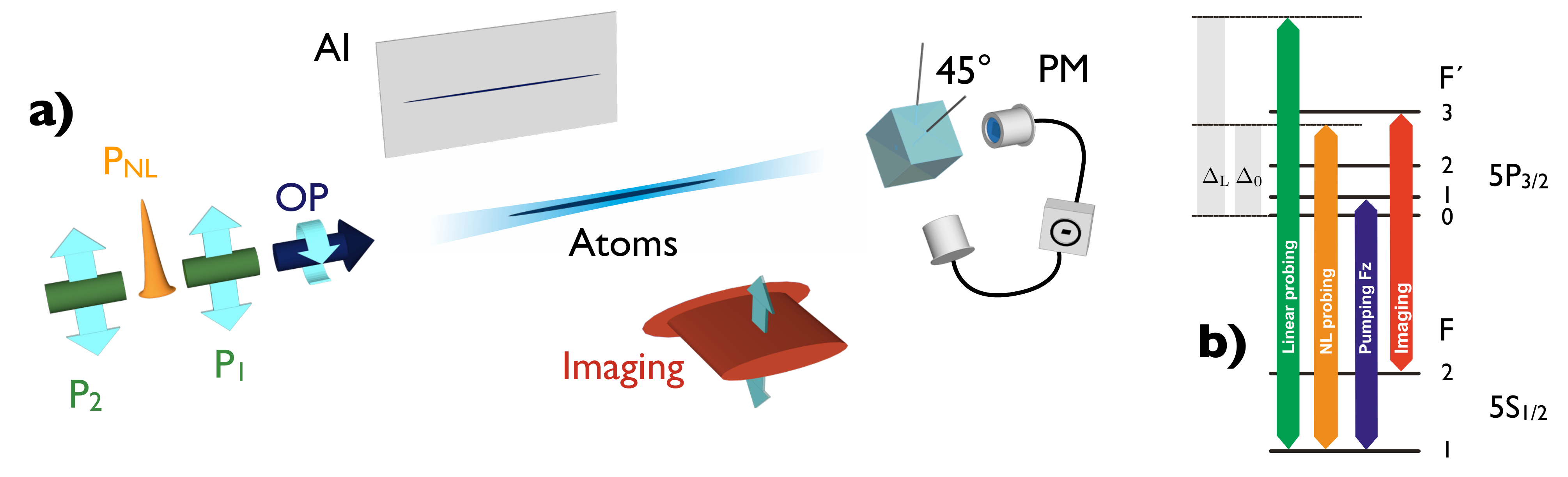}
    \caption{
    \textbf{Atom-light interface} a) Experimental schematic: an
ensemble of  \( 7 \times 10^5 \) \rb\ atoms, held in an optical
dipole trap, is prepared in \(
\left|F=1,m_F=1\right> \) by optical pumping (OP).  Linear
(P$_1$), nonlinear (P$_{\rm NL}$), and a second linear (P$_2$)
Faraday rotation probe pulses measure the atomic magnetisation,
detected by a shot-noise-limited polarimeter (PM). The atom number
is measured by quantitative absorption imaging (AI). b) Spectral
positions of the pumping, probing, and imaging light on the D$_2$
    transition.
    }
    \label{fig:setup}
\end{figure}

In this Letter, we study interaction-based quantum
metrology using unentangled probe particles. One challenge in
demonstrating SH scaling is to engineer a suitable nonlinear
Hamiltonian.  Some nonlinearities have been shown to be
intrinsically noisy\cite{FleischhauerPRA2000} while others give SH
scaling but fall short of the ideal $N^{-(k-1/2)}$ under realistic
conditions\cite{BoixoPRL2008,TaclaPRA2009}.  We use a cold atomic
ensemble as a light-matter quantum interface\cite{HammererRMP2010}
to produce quantum-noise-limited interactions and a Hamiltonian of
the form $\OpH =\hbar \rotstr \OpS_z \OpS_0 =\hbar \rotstr \OpS_z
N/2$. This Hamiltonian gives a polarisation rotation growing with
the photon number, without increasing quantum
noise\cite{BoixoPRL2008}.
The experiment, shown schematically in Fig.~\ref{fig:setup}, uses
pulses of near-resonant light to measure the
collective spin $\hat{\bf F}$ of an ensemble of $N_A  \sim 10^6$
cold rubidium-87 atoms, probed on the $5S_{1/2}\rightarrow
5P_{3/2}$ D$_2$ line.
The experimental system is described in detail in the
references\cite{KubasikPRA2009,KoschorreckPRL2010a}. The on-axis
atomic magnetisation \( \myexpect{\OpF_z} \), which plays the role
of $\unknown$ in this measurement, is prepared in the initial
state \( \myexpect{\OpF_z}= N_A\) by optical pumping with resonant
circularly polarised light propagating along the trap axis $z$. A weak on-axis magnetic field is applied to
preserve \( \OpF_z \) during the measurements.

Pulses of $\OpS_x$ polarised, but not entangled, photons pass through
the ensemble and experience an optical rotation
proportional to $\myexpect{\OpF_z}$. The light-atom
interaction Hamiltonian \( \OpH_{\rm eff} = \alpha^{(1)} \OpF_z
\OpS_z +  \beta^{(1)} \OpF_z \OpS_z N/2\) describes this
paramagnetic Faraday rotation\cite{NapolitanoJP2010}.
 Both the linear term $\alpha^{(1)}
\OpF_z \OpS_z$ and the nonlinear term $ \beta^{(1)} \OpF_z
\OpS_z N/2$ cause rotation of the plane of polarisation from
$\OpS_x$ (vertical) toward $\OpS_y$ (diagonal).  Detection of
$\OpS_y$ then allows estimation of $\OpF_z$. As described
in the Supplementary Information, $\alpha^{(1)}$ and $\beta^{(1)}$
depend on the optical detuning $\Delta$ relative to the $F=1
\rightarrow F'=0$ transition, in particular $\alpha^{(1)}(\Delta_0) =0$ for
the specific detuning $\Delta_0 \approx 2\pi(\SI{468.5}{MHz})$, allowing
a purely nonlinear estimation to be studied.


 The rotation angle is \( \phi=\myexpect{\OpF_z}
[A(\Delta)+B(\Delta) \NumGen]/2  \) where $A \propto \alpha^{(1)}$
and $B \propto \beta^{(1)}$ account for the temporal pulse shape
and geometric overlap between the atomic density and the spatial
mode of the probe. The shot-noise limited uncertainty in the
rotation angle, due to quantum uncertainty in the initial angle,
is $\delta\phi =
N^{-1/2}/2$.  A contribution $\myexpect{\OpF_z} B(\Delta)
\delta\NumGen/2$ from initial number fluctuations $\delta\NumGen =
\expect{N}^{-1/2}$ is negligible for small rotation angles.  This
gives a measurement uncertainty
\begin{equation}
\label{eqn:SH-scaling}
    {\delta F_z} = {\myexpect{\OpF_z}}\frac{\delta \phi}{\phi}
    = \frac{1}{A(\Delta) \NumGen^{1/2}+B(\Delta) \NumGen^{3/2}},
\end{equation}
indicating a transition from  SQL scaling \( \delta F_z
\propto \NumGen^{-1/2} \) to
SH scaling \( \delta F_z \propto \NumGen^{-3/2} \)
with increasing $\NumGen$.

Two regimes of probing are used:  the {\it{linear probe}}
consists of forty \SI{1}{\micro s} pulses (total illumination time
$\tau_{\rm L}=\SI{40}{\micro s}$) spread over \SI{400}{\micro s}
with detuning $\Delta_{\rm L}\gg \Delta_0$. This gives $A\gg
N_{\rm L} B$, i.e., linear estimation and, as  described by
Koschorreck {\it et al.}\cite{KoschorreckPRL2010a}, provides a
projection-noise-limited quantum-non-demolition (QND)
measurement\cite{BraginskiSPU1975} of \( \OpF_z \), with
uncertainty at the parts-per-thousand
level\cite{KoschorreckPRL2010a}. The {\it{nonlinear probe}}
consists of a single $\tau_{\rm NL} = \SI{54}{\ns}$ FWHM,
Gaussian-shaped, high-intensity pulse with $\NumPhot$ photons and
detuning $\Delta_0$, so that $A\ll \NumPhot B$. Crucially, having
two probes allows us to precisely calibrate the nonlinear
measurement using a highly sensitive and well characterised
independent measurement of the same sample.

We probe the same sample three times for each preparation:
First with the linear probe, which gives a precise and
non-destructive measurement of $\myexpect{\OpF_z}$ via a rotation
\( \phi_{\rm{L}} \). Then with the nonlinear probe, contributing
with a rotation \( \phi_{\rm{NL}} \), which is calibrated against
the ``true'' value (i.e., with negligible error) provided by the
previous linear probe. Third, a second linear probe is used to
estimate the damage to the atomic magnetisation $\eta \equiv
1-\phi_{\rm{L}'}/\phi_{\rm L}$ caused by the nonlinear probe.

%

\begin{figure}[h]
    \centering
   \includegraphics[width=.75\textwidth]{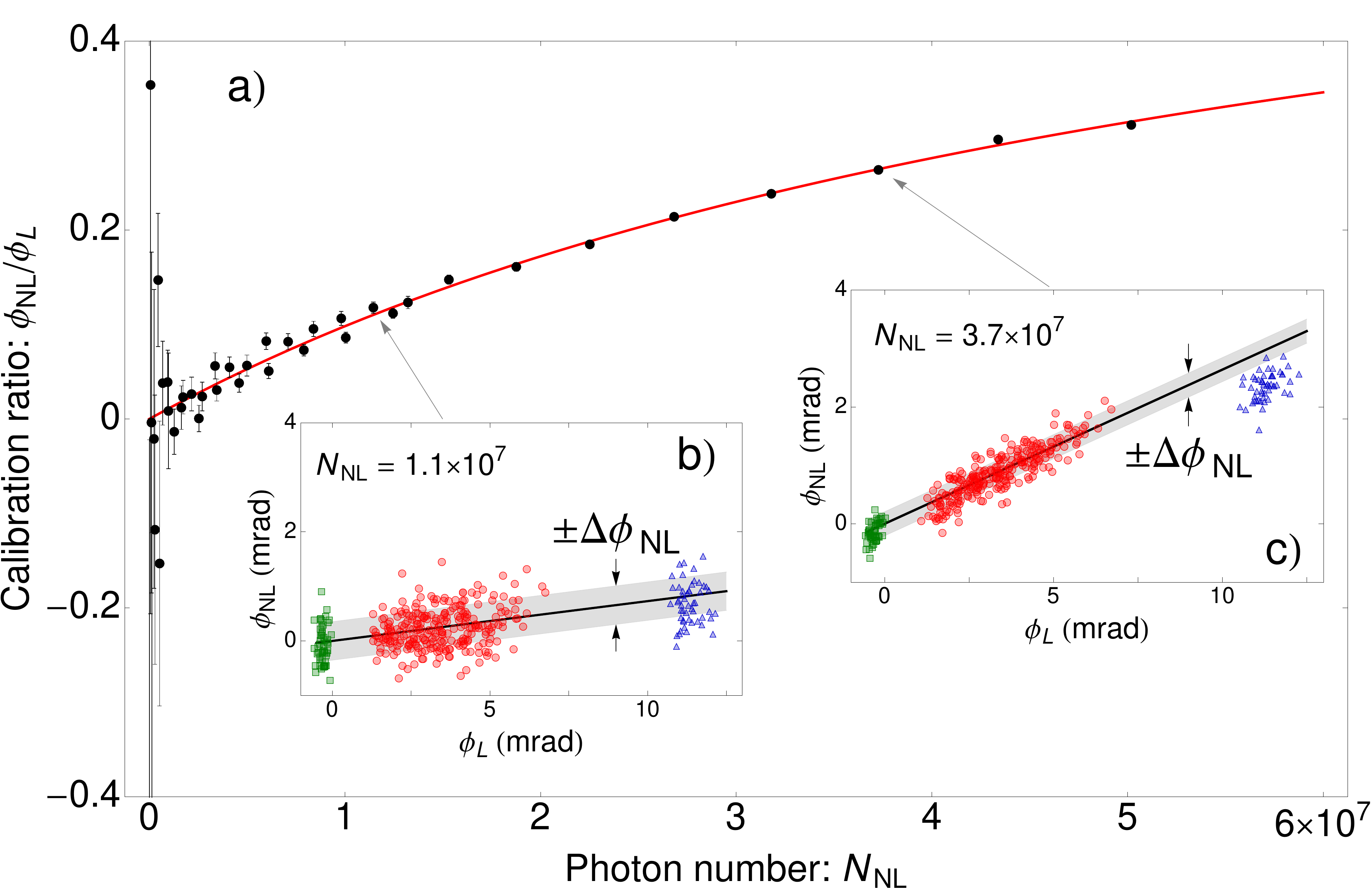}
    \caption{
 \textbf{Calibration of nonlinear Faraday rotation.} a) Ratio of nonlinear rotation $\phi_{\rm{NL}}$ to linear rotation $\phi_{\rm{L}}$ {\it vs.} nonlinear probe photon number \( \NumPhot \). The data points and error bars indicate best fit and standard errors from a linear regression \( \phinl = b\philin + {\rm const.} \) for a given $\NumPhot$. The red curve is a fit with equation (\ref{eqn:saturation}), showing the expected nonlinear behaviour $\phi_{\rm NL} \propto \NumPhot$, with some saturation for large $\NumPhot$. b) \& c)  \corrplots~for two values of \( \NumPhot \).  The atom number \( N_A \) is varied to produce a range of $\phi_{\rm L}$ and $\phi_{\rm NL}$. Green squares: no atoms $N_A=0$, red circles: $1.5\times 10^5 < N_A < 3.5 \times 10^5$, blue triangles $N_A \approx 7 \times 10^5$.  The blue triangles are shown as a check on detector saturation, and are not included in the analysis.
 }
    \label{fig:calibration}
\end{figure}

The linear probe is calibrated using quantitative
absorption imaging to measure $N_A$, and we find  \( A(\Delta_{\rm
L}) = 3.3(1)\times10^{-8}\si{rad} \) per atom. The calibration of
the nonlinear probe against the first linear probe is shown in
Fig.~\ref{fig:calibration}:  We repeat the above pump/probe
sequence while varying \( N_A \) in the range $N_A = 1.5 \times
10^5$ to $N_A = 3.5 \times 10^5$ to generate a \corrplot~for a
given $\NumPhot$.  Since both $\philin$ and $\phinl$ are linear in
$N_A$, we use linear regression to find the slope $b = d \phinl /
d \philin = {B(\Delta_0)\NumPhot}/{A(\Delta_{\rm L})}$ for that
value of \( \NumPhot \). The experiment is repeated varying the
number of photons \( \NumPhot \) in the nonlinear pulse.


The observed \( b \) {\it vs.} \( \NumPhot \), shown in
Fig.~\ref{fig:calibration}a, is well fit by a simple model
including saturation of the nonlinear response:

\begin{equation}
\label{eqn:saturation}
    \frac{d\phinl}{d\philin}= \frac{B(\Delta_0)\NumPhot}{A(\Delta_L)} \frac{1}{1+\NumPhot/\NumPhot^{(\rm{sat})}},
\end{equation}

\noindent  with a saturation parameter \( \NumPhot^{(\rm{sat})} =
6.0(8) \times 10^7 \) and the nonlinear coupling strength
\(B(\Delta_0) = 3.8(2) \times 10^{-16} \si{rad}\) per atom per
photon.

\begin{figure}[h]
    \centering
   \includegraphics[width=.8\textwidth]{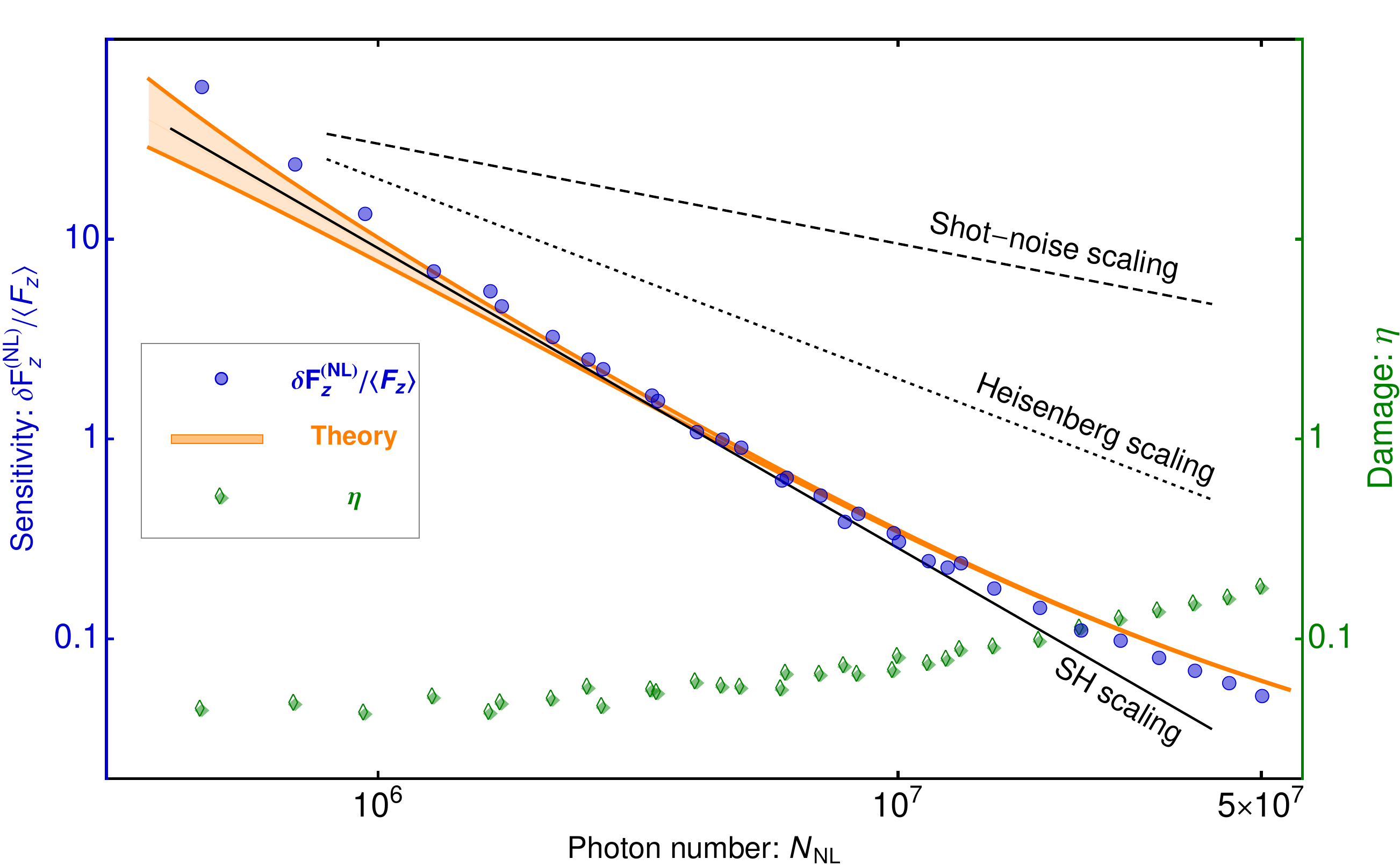}
    \caption{
    \textbf{Super-Heisenberg scaling.} Fractional
sensitivity \(\delta F_z^{(\rm
NL)}/\myexpect{\OpF_z} \) of the nonlinear probe versus number of
interacting photons $\NumPhot$. Blue circles indicate the measured
sensitivity, curves show results of numerical modelling, and the
black lines indicate SQL, HL, and SH scaling for reference.
Scaling surpassing the Heisenberg limit $\propto \NumPhot^{-1}$ is
observed over two orders of magnitude.  The measured damage $\eta$
to the magnetisation, shown as green diamonds, confirms the
non-destructive nature of the measurement. {Error bars for
standard errors would be smaller than the symbols and are not
shown.}
}
    \label{fig:scaling}
\end{figure}

The noise in the nonlinear probe, again as a function of
$\NumPhot$, is determined from the \corrplots.  As illustrated in
Fig.~\ref{fig:calibration}b-c, the residual standard deviation of
the fits indicates the observed uncertainty $\Delta
\phi_{\rm{NL}}$, which includes the intrinsic uncertainty $\delta
\phi_{\rm{NL}}$ and a small contribution from electronic noise.
In Fig.~\ref{fig:scaling} we plot the
fractional sensitivity \( \delta F_z^{(\rm NL)}/\myexpect{\OpF_z}
\) {\it vs.} \( \NumPhot \), calculated using equation
(\ref{eqn:saturation}) and considering the whole polarised
ensemble, $\myexpect{\OpF_z}=7 \times 10^5$.  In agreement with
equation (\ref{eqn:SH-scaling}), the log-log slope indicates the
scaling \( \delta F_z^{(\rm NL)} \propto \NumPhot^{-3/2} \) to
within experimental uncertainties in the range $\NumPhot = 10^6$
to $\NumPhot = 10^7$, and SH scaling, i.e.{,} steeper than
$N^{-1}$, over two orders of magnitude $\NumPhot = 5 \times 10^5$
to $\NumPhot = 5 \times 10^7$.

Results of numerical modelling using the Maxwell-Bloch equations
to describe the nonlinear light-atom interaction are also shown in
Fig.~\ref{fig:scaling}.  Two curves are shown, for detunings
$\Delta_0 \pm 2 \pi (\SI{200}{kHz})$, covering the combined
uncertainty in $\Delta$ due to the probe laser linewidth and
inhomogeneous light shifts in the optical dipole trap.  As
expected from equation (\ref{eqn:SH-scaling}), this alters the
sensitivity only at low $\NumPhot$. {The model is described in
detail in the Supplementary Information.}

For photon numbers above \( \NumPhot \gtrsim 2 \times 10^7 \),
the saturation of the nonlinear rotation alters the slope. This
can be understood as optical pumping of atoms into states other
than  \( \left|F = 1, m_F =1\right> \) by the nonlinear probe. The
damage to the atomic magnetisation \(
\eta=1-\phi_{\rm{L}'}/\phi_{\rm{L}} \), shown in
Fig.~\ref{fig:scaling} remains small, confirming the
non-destructive nature of the measurement. The finite damage even for small \(
\NumPhot \) is possibly due to stray light and/or magnetic fields
disturbing the atoms during the \SI{20}{ms } period between the
two linear measurements. At large $N$,
high-order nonlinear effects including optical pumping limit the
range of SH scaling.


The experimental results 
illustrate the subtle
relationship of scaling to sensitivity in a nonlinear system. For
an ideal nonlinear measurement, the improved scaling would
guarantee better absolute sensitivity for sufficiently large $N$.
%
%
%
%
%
%
Indeed, when the measurement bandwidth is taken into account, the nonlinear probe overtakes the linear one at $N = 3.2\times 10^6$ where both achieve a sensitivity of $ 1.1\times 10^{2} ~{\rm spins}~\SI{}{\PerRtHz}$. As a consequence, the nonlinear technique performs better in fast measurements.  In contrast, when measurement time is not a limited
resource, the comparison can be made on a ``sensitivity-per-measurement'' basis, and the ideal crossover point  of $3.2\times 10^3 ~{\rm spins}$ at $N = 8.7\times 10^7$ is never actually reached, due to the higher-order nonlinearities.
Evidently SH scaling enables but does not guarantee enhanced
sensitivity: for the nonlinear to overtake the linear, it is also
necessary that the scaling extend to large enough $N$. The
comparison shows also that resource constraints dramatically
influence
 the linear vs. nonlinear comparison. See also the
Supplementary Information.

We have realised a scenario proposed by Boixo {\it et
al.}\cite{BoixoPRL2007} to achieve metrological sensitivity beyond
the Heisenberg limit $\delta \unknown \propto N^{-1}$ using
metrologically-relevant interactions among particles.  To generate
pairwise photon-photon interactions, we use fast nonlinear optical
effects in a cold atomic ensemble and measure the ensemble
magnetisation $\myexpect{\OpF_z}$ with super-Heisenberg sensitivity $\delta
F_z \propto N^{-3/2}$.  To rigorously quantify the
photon-photon interaction and the sensitivity, we calibrate
against a precise, non-destructive, linear measurement of the same
atomic quantity\cite{KoschorreckPRL2010a}, demonstrate
quantum-noise-limited  performance of the optical instrumentation,
and place an upper limit on systematic, i.e., non-atomic,
nonlinearities at the few-percent level. The experiment demonstrates the use of inter-particle
interactions as a new resource for quantum metrology. While
possible applications to precision measurement will require
detailed study, this first experiment shows that interactions can
produce super-Heisenberg scaling and improved precision in a
quantum-limited measurement.

\begin{methods}
\label{sec:methods}

\subsection{Linear \& nonlinear probe light.}

The probe beam is aligned to the axis of the trap with a waist of
\SI{20}{\micro m}, chosen to match the radial dimension of the
cloud. In the linear probing regime we use a train of forty
\SI{1}{\micro s} pulses, pulse period \SI{10}{\micro s}, each
containing \( 3 \times 10^6 \) photons detuned \( +\SI{1.5}{GHz}
\) from the (\( F=1\rightarrow F'=0 \)) transition. The maximum
intensity is \SI{0.1}{W cm^{-2}}. The signals are summed and can
be considered a single, modulated pulse.

The nonlinear probe consists of a single Gaussian-shaped pulse
with a FWHM of \SI{54}{\ns}. The maximum intensity of the
nonlinear probe is \SI{7}{W cm^{-2}} for a pulse with \( 10^7 \)
photons. Theory predicts \( \alpha^{(1)}=0 \) at a detuning \(
\Delta = 2\pi(\SI{462}{MHz}) \) in free space. This is modified by
trap-induced light shifts, and we use the empirical value \(
\Delta_0 = 2\pi(\SI{468.5}{MHz}) \), which gives zero rotation at
low probe intensity.

\subsection{Instrumental noise.}

The instrumental noise is quantified by measuring  $\var(\OpS_y)$ {\it vs.} input photon number $N$ ($=N_L$ or $\NumPhot$), in the absence of atoms, to find contributions from electronic noise $V^{(\rm el)}\propto N^0$, shot-noise $= N^1$, and technical noise $\propto N^2$, as described in the Supplementary Information. We find $V_{\rm L}^{(\rm el)}$ and $V_{\rm NL}^{(\rm el)}$ are \( 3 \times 10^5 \) and   \( 4\times 10^5 \)  per pulse, respectively, while the technical noise is negligible.  The instrumentation is thus shot-noise-limited over the full range of $N$ used in the experiment. The intrinsic rotation uncertainty of the nonlinear probe $\delta\phi_{\rm NL}$ is calculated from the measured $\dphiexpt$ as $(\delta\phi_{\rm NL})^2 = (\dphiexpt)^2 - V_{\rm NL}^{(\rm el)}$.  The correction is at most 5\%.

{\subsection{Instrumental linearity.} The linearity of the
experimental system and analysis is verified using a wave-plate in
place of the atoms to produce a linear rotation equal to the
largest observed nonlinear rotation.  Over the full range of
photon numbers used in the experiment, the detected rotation angle
is constant to within $5\%$, and SQL scaling is observed.}
\end{methods}




\begin{addendum}
\item We thank I. H. Deutsch and F. Illuminati, for helpful comments. We thank C.~M.~Caves and A. D. Codorn\'iu for inspiration.
{This work was supported by the Spanish Ministry of Science
 and Innovation through the Consolider-Ingenio 2010 project QOIT,
Ingenio-Explora project OCHO (Ref. FIS2009-07676-E/FIS), project
ILUMA (Ref. FIS2008-01051) and by the Marie-Curie RTN EMALI.}

 {
 \item[Author Contributions] All authors contributed equally to the work presented in this paper.}

 \item[Competing Interests] The authors declare that they have no
competing financial interests.
 \item[Correspondence] Correspondence and requests for materials
should be addressed to Mario Napolitano~(email: mario.napolitano@icfo.es).
\end{addendum}

\newpage

\newpage

\newpage

\newpage

\newcommand{\captionfonts}{\footnotesize}

\makeatletter  
\long\def\@makecaption#1#2{%
  \vskip\abovecaptionskip
  \sbox\@tempboxa{{\captionfonts #1: #2}}%
  \ifdim \wd\@tempboxa >\hsize
    {\captionfonts #1: #2\par}
  \else
    \hbox to\hsize{\hfil\box\@tempboxa\hfil}%
  \fi
  \vskip\belowcaptionskip}
\makeatother   

\renewcommand{\figurename}{{\bf{Supp. Fig.}}}

\newcommand{\myvec}[1]{{\bf #1}}
\newcommand{\myvecE}[1]{\pmb{#1}}

\newcommand{\OpJ}{{\hat{J}}}
\newcommand{\Opf}{{\hat{f}}}
\newcommand{\Opj}{{\hat{j}}}
\newcommand{\bff}{{\bf f}}
\setcounter{figure}{0}

\noindent
{\Large{\bfseries\sffamily Supplementary Information for ``Interaction-based quantum metrology showing scaling beyond the Heisenberg limit.''}}

\noindent M. Napolitano$^{1}$, M. Koschorreck$^1$,
B. Dubost$^{1,2}$, N. Behbood$^1$, R. J. Sewell$^1$ \& M. W. Mitchell$^1$

\begin{affiliations}
 \item ICFO-Institut de Ciencies Fotoniques, 08860 Castelldefels (Barcelona), Spain.
 \item Laboratoire Mat\'{e}riaux et Ph\'{e}nom\`{e}nes Quantiques, Universit\'{e} Paris Diderot et CNRS, \\UMR 7162, B\^{a}t. Condorcet, 75205 Paris Cedex 13, France.
\end{affiliations}

\section{Supplementary Discussion}

\subsection{Interaction Hamiltonian.}

The atom-light interaction is described using collective continuous variables and
degenerate perturbation theory in reference [13] of the main text. 
We repeat essential results:

The electric dipole interaction $h_{\rm int} = - \bE\cdot \myvec{d}$, taken in second order perturbation theory, gives rise to an effective (single-atom) Hamiltonian of the form
\begin{equation}
\tag{S1}
H_{\rm eff}^{(2)}  =   \alpha^{(1)}\OpS_z\OpJ_z+ \alpha^{(2)}\left(\OpS_x\OpJ_x+\OpS_y\OpJ_y\right),\label{eq:HEff2}
\end{equation}
plus terms in $\OpS_0$ which do not alter the optical polarisation.  Here $\alpha^{(1)},\alpha^{(2)}$ describe the vectorial and tensorial components of the interaction respectively, and the atomic collective variable is $\hat{\bJ} \equiv \sum_i \hat{\bj}^{(i)}$ where the superscript $(i)$ indicates the $i$'th atom  and  $\Opj_x \equiv \left(\Opf_x^{2}-\Opf_y^{2}\right)/2$, $\Opj_y \equiv \left(\Opf_x \Opf_y+\Opf_y \Opf_x\right)/2$, $\Opj_z \equiv \Opf_z/2$ and $\Opj_0 \equiv \Opf_z^2/2$. {For our case of the F=1 ground state, the $\hat{\bj}$ operators, defined starting from the angular momentum operators $\hat{\bff}$, represent a pseudo-spin 1/2 system involving the states $m_f=\pm 1$. In this representation, the measurement of $\OpF_z$, described in the main text, is equally to a measurement of $2 \OpJ_z=\OpF_z$.}


\begin{figure}
  \begin{center}
    \begin{minipage}[t]{0.58\linewidth}
      \includegraphics[width=1\textwidth]{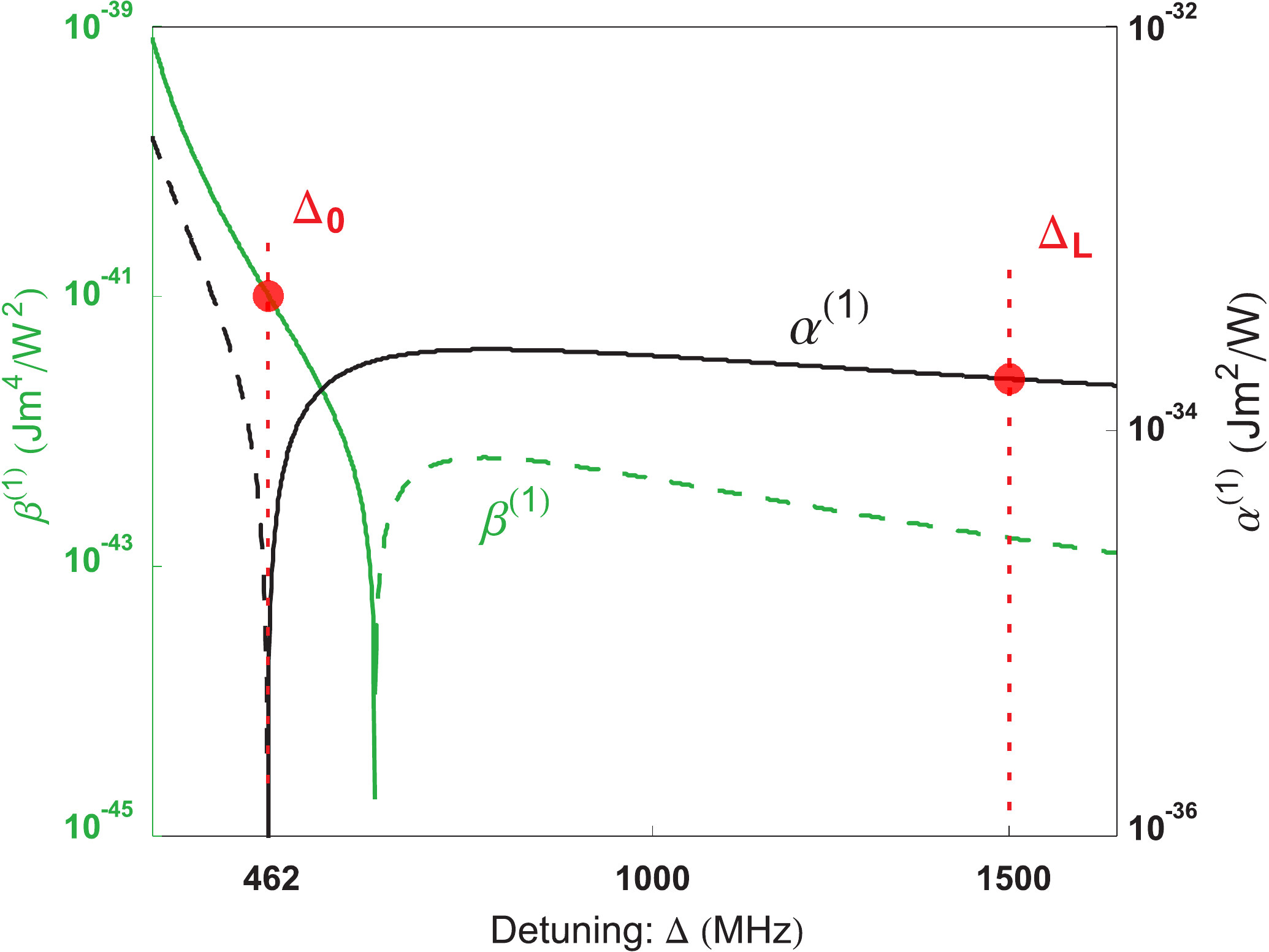}
    \end{minipage}\hfill
    \begin{minipage}[b]{0.38\linewidth}
      \caption{\textbf{Spectra of the terms of the effective Hamiltonian.} Black curve and right axis correspond to $\alpha^{(1)}$; Green curve and left axis to $\beta^{(1)}$. The curves are dashed when the coefficient assumes negative value. Detuning [MHz] is relative to the $F=1\rightarrow F^\prime=0$ of ${}^{87}$Rb $D_2$ transition.  The detunings of the two probing regimes used in the experiment are indicated.}
\label{fig:alphabeta}
    \end{minipage}
  \end{center}
\end{figure}

For $F=1$ atoms, the fourth-order contribution (again ignoring terms depending only on $\OpS_0$) is:
\begin{equation}
\tag{S2}
H_{\rm eff}^{(4)} = \beta_J^{(0)} \OpS_z^2 \OpJ_0 + \beta_N^{(0)}
\OpS_z^2 N_A +  \beta^{(1)}  \OpS_0 \OpS_z \OpJ_z
+
\beta^{(2)}  \OpS_0 (\OpS_x \OpJ_x + \OpS_y \OpJ_y).
\end{equation}
For our input state, consisting of $N$ vertically-polarized photons, i.e.,
$\left<\right. (\OpS_x,\OpS_y,\OpS_z) \left.\right> = (N/2,0,0)$, we can drop all but the
$\alpha^{(1)}$ and $\beta^{(1)}$ terms, because 1)
terms in $\OpS_x$ and $\OpS_0\OpS_x$, leave the initial
state unchanged, 2) terms in $\OpS_y$ and $\OpS_0 \OpS_y$ commute with the measured variable,
giving no measurable signal and 3) the terms in $\OpS_z^2$ make a contributions smaller than the
$\beta^{(1)}$ term by a  factor $\sim \OpS_z/\OpS_0$.

%
%
%

The coefficients $\alpha^{(1)}$ and $\beta^{(1)}$ depend strongly on the probe frequency due to the excited state hyperfine structure.
For the $D_2$ line of $^{87}$Rb, from the $F=1$ ground state, they are shown graphically in Supp. Fig. 1.

{We note also that $H_{\rm eff}^{(4)}$ is sensitive to more spin
degrees of freedom than is $H_{\rm eff}^{(2)}$.  The population of
the state $\ket{F=1,m=0}$, i.e., $N_A-J_0$, appears in $H_{\rm
eff}^{(4)}$ proportional to $\OpS_z^2$ and produces polarization
self-rotation. In contrast, $H_{\rm eff}^{(2)}$ has no dependence
on this population, which cannot be detected by any linear
measurement.  }

\subsection{Atomic State Preparation.}

The atomic ensemble contains up to \( 7 \times 10^5 \) \rb\  atoms held in an optical dipole trap formed by a weakly-focused (\SI{52}{\micro m}) beam of a Yb:YAG laser at \SI{1030}{nm} with \SI{6}{W} of optical power. The trap is loaded from a conventional magneto-optical trap (MOT) and cooled to  \SI{25}{\micro K} with sub-Doppler cooling. The system has demonstrated high effective optical depth$^{23}$  (\( d_0>50 \)) and sub-projection-noise sensitivity$^{8}$ of $\sim$ 500 spins with the linear probe.

The ensemble is polarised
\( \myexpect{\OpF_z} = N_A \)  by optical pumping
with circularly polarised light resonant with the \( F=1\rightarrow
F'=1 \) transition, sent along the longitudinal axis of the trap.   Repump light resonant with \( F=2\rightarrow F'=2 \) is
simultaneously applied via the 6 directions of the MOT beams
to prevent accumulation in the \( F=2 \) hyperfine level.  A small
bias magnetic field of \SI{100}{m G} is applied along the axis to preserve $\OpF_z$.

Before each polarisation step, the state of the ensemble is reset to a fully-mixed state by repeated pumping from $F=1$ to $F=2$ and back, using  resonant lasers from the MOT beams as described in Koschorreck {\it et al.}$^{8}$.  During the reset process, about 10\% of the atoms escape from the trap, allowing measurement with different \( N_A \) during a single loading cycle.

\begin{figure}
  \begin{center}
    \begin{minipage}[t]{0.58\linewidth}
      \includegraphics[width=1\textwidth]{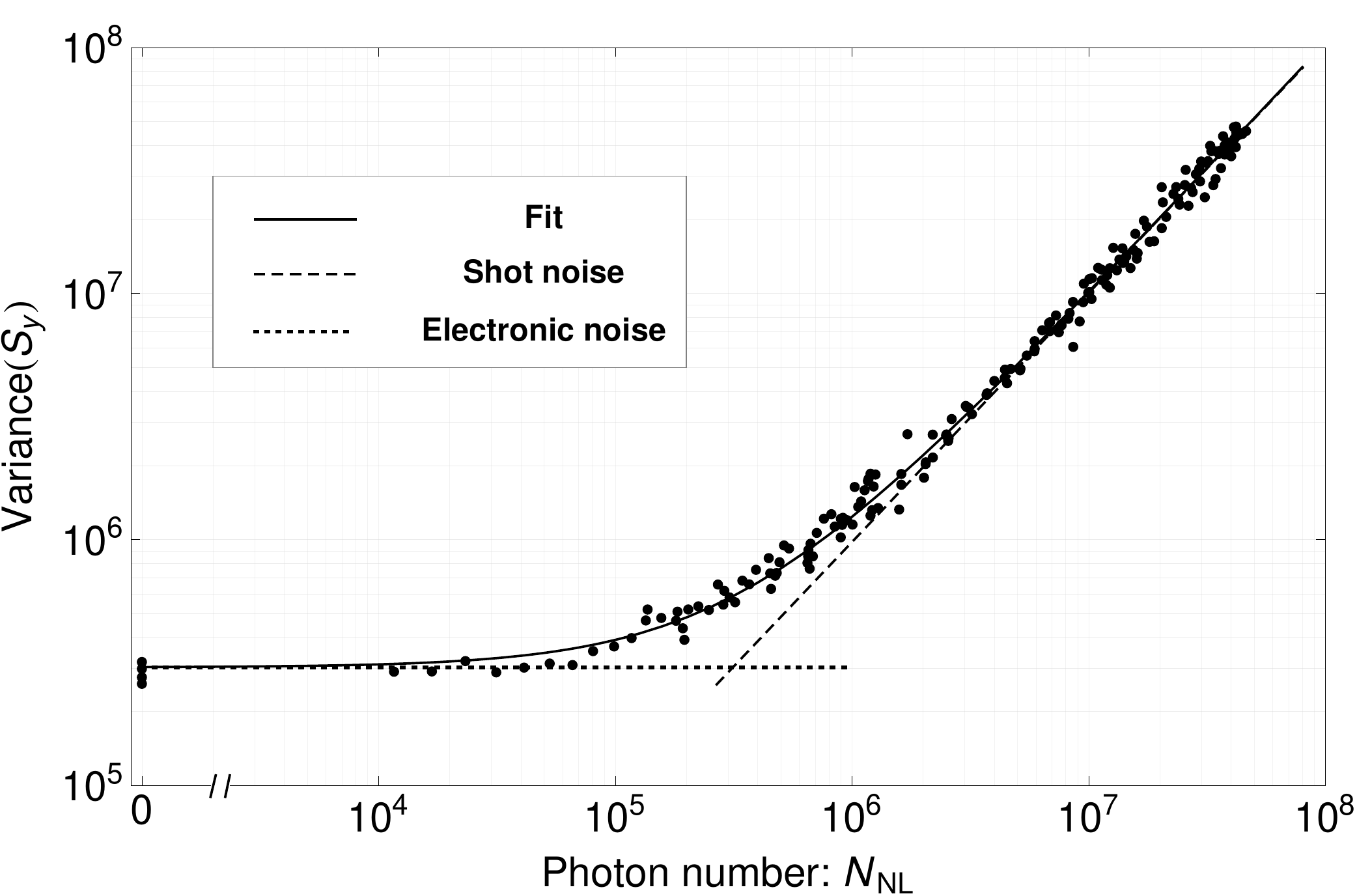}
    \end{minipage}\hfill
    \begin{minipage}[b]{0.38\linewidth}
     \caption{\textbf{Shot noise limited detection.}
    {Detected variance in $\OpS_y$ versus input photon number for short, intense pulses of the nonlinear probe.  Points are measured variances, curve is a fit to $\var(\OpS_y) = V_{\rm el} +  \NumPhot$, where $V_{\rm el}$ is the electronic noise.  Negligible technical noise is seen. The system is shot-noise limited above  \( \NumPhot \gtrsim 4 \times 10^5 \), i.e., for the full range used in the experiment.   }
   }
\label{fig:shotnoise}
    \end{minipage}
  \end{center}
\end{figure}

\subsection{Shot noise limited detection.}
{Before the ensemble, a  beamsplitter and {calibrated} fast
photodiode are used to detect the input pulse energy  $\OpS_0 =
\OpS_x$.
After the ensemble, pulses are analyzed in the \( \pm\ang{45} \) basis with an ultra-low-noise balanced photo-detector$^{25}$, 
giving a direct measure of \( \OpS_y \).  Both signals are
recorded on a digital storage oscilloscope, and rotation angles
calculated as $\phi = \OpS_y/(\OpS_x \sqrt{T_H T_V})$ where
$T_{H,V}$ are the measured transmission coefficients for the
system optics (vacuum cell, lenses and dichroic mirrors to
separate the dipole trap beam).  Supp. Fig. \ref{fig:shotnoise}
shows the noise {\it vs.} power curve for generation and detection
of nonlinear probe pulses, indicating an electronic noise
contribution to $\var(\OpS_y)$ of \( 4\times 10^5 \) per pulse.
This electronic noise is subtracted when calculating $\delta
\phi_{\rm{NL}}$ in Fig.  3 of the main text.}

\begin{figure}
  \begin{center}
    \begin{minipage}[t]{0.58\linewidth}
      \includegraphics[width=1\textwidth]{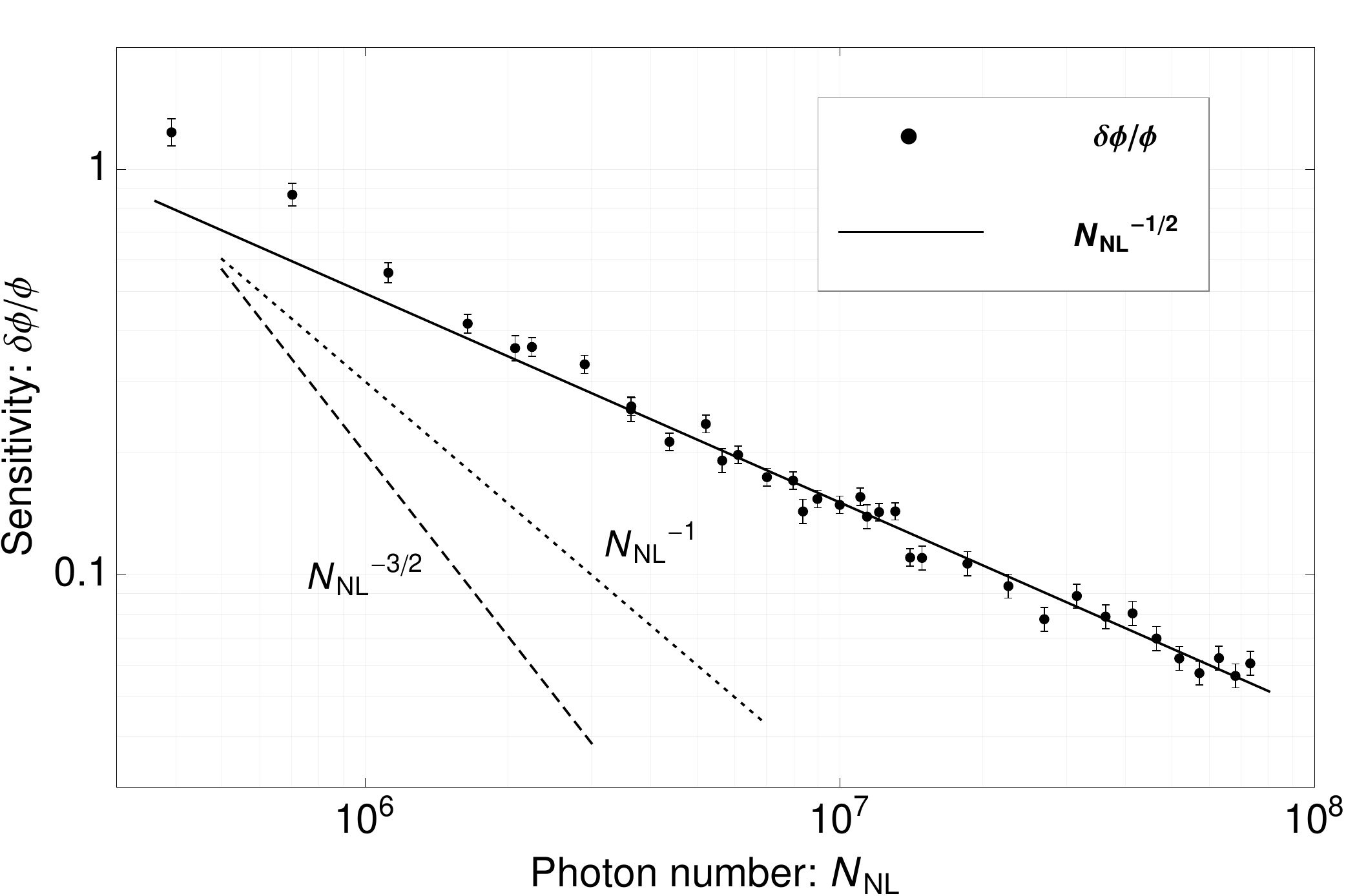}
    \end{minipage}\hfill
    \begin{minipage}[b]{0.38\linewidth}
    \caption{\textbf{ {System linearity}.}
{Sensitivity results obtained as in Fig. 3 of the main text, but
with a waveplate rotation in place of the atomic Faraday rotation.
As expected the sensitivity shows SQL scaling, providing a direct
verification of the linearity of the equipment and method of
analysis. Error bars plotted are the standard errors of the
measured rotation
     signal.}
     }\label{fig:reference}
    \end{minipage}
  \end{center}
\end{figure}


\subsection{System linearity.}
{To check for systematic errors, both in the apparatus and in the
analysis, we repeat the experiment under identical conditions but
with no atoms present.  We mimic the Faraday rotation signal by
rotating the wave-plate used to balance the polarimeter to give a
signal equal to the largest signal seen with atoms. The measured
rotation is independent of $\NumPhot$, and gives shot-noise
scaling of the sensitivity, plotted in Supp. Fig.
(\ref{fig:reference}) over the range of $\NumPhot$ used in the
experiment.}

\subsection{Modelling.}

We model the nonlinear rotation by integrating the Maxwell-Bloch
equations in three spatial dimensions $\bx = (x,y,z)$ plus time
$t$.  This semiclassical model describes the average rotation
$\myexpect{\phi} = \myexpect{\OpS_y^{({\rm
out})}}/\myexpect{\OpS_x^{({\rm in})}}$, which remains $\ll 1$,
while the quantum noise is given by $\delta \phi = \delta
\OpS_y^{({\rm out})}/\myexpect{\OpS_x^{({\rm in})}} \approx \delta
\OpS_y^{({\rm in})}/N  = 1/\sqrt{N}$.

\newcommand{\DownwardDipole}{\myvec{d}_\DownSym}
\newcommand{\UpwardDipole}{\myvec{d}_\UpSym}

\renewcommand{\cE}{{{\cal E}}}
\renewcommand{\cP}{{{\cal P}}}
\newcommand{\cEVec}{{\myvecE{\cE}}}
\newcommand{\cPVec}{{\myvecE{\cP}}}
\newcommand{\EIn}{{{\cE}^{({\rm in})}}}
\newcommand{\EInPar}{{{\cE}^{({\rm in})}_V}}
\newcommand{\EInVec}{{\myvecE{\cE}^{({\rm in})}}}
\newcommand{\beamshape}{M}
\newcommand{\pulseshape}{T}
\newcommand{\overlap}{{\cal O}}
\newcommand{\SigPerAtom}{{\cal Q}}
\newcommand{\InputStr}{{\cal E}_0}
\newcommand{\InputPol}{{\myvec{e}}_{\rm in}}
\newcommand{\PolV}{{\myvec{e}}_{V}}
\newcommand{\OutputPol}{{\bf e}_{H}}
\newcommand{\OneAtomPolVec}{\myvec{p}}
\newcommand{\OneAtomPol}{{p}}
\newcommand{\Supplement}{Methods section}

In retarded coordinates $\zeta \equiv z $ and $\tau \equiv t -
z/c$, the field envelope $\cEVec(\bx,\tau)$ and atomic state
$\rho(\bx,\tau)$  {obey the coupled} equations
\begin{align*}
\label{Eq:PWEcm}\tag{S3}
{\cal D} \cEVec &= \frac{k^2}{ \varepsilon_0} {\cPVec}
  \\
\label{Eq:oBEcm}
 \partial_\tau \rho &= \frac{i}{\hbar}  [\rho,H(\cEVec)] + {\cal L}(\rho)\tag{S4}\end{align*}
 where ${\cal D} \equiv \partial_x^2 + \partial_y^2+ 2 i
k \partial_\zeta $ is the differential operator of the paraxial
wave equation (PWE), $k$ is the wave-number,
${\cal L}$ is the Liouvillian describing relaxation  and
{the polarization envelope $\cPVec(\bx,\tau)$ is } \begin{align*}\tag{S5}
\label{Eq:PBEcm} {\cPVec} \equiv n{\rm Tr}[\rho
\DownwardDipole\, ] \equiv n \OneAtomPolVec,\end{align*}  where $n$ is the
local atomic number density and  $\DownwardDipole$ is the dipole
operator describing downward transitions.  For the atom
distribution, we take a Gaussian with FWHM $2 \sigma_T \sqrt{\ln
2}$ and $2 \sigma_L \sqrt{\ln 2}$ in the transverse and longitudinal
directions, respectively:
 \begin{align*}\tag{S6} n(\bx) = N_A
(\pi^{3/2} \sigma_L \sigma_T^2)^{-1} \exp[-r^2/\sigma_T^2]
\exp[-z^2/\sigma_L^2]\label{Eq:GaussDens} \end{align*} where $r^2\equiv x^2+y^2$.

\begin{figure}
  \begin{center}
    \begin{minipage}[t]{0.58\linewidth}
      \includegraphics[width=1\textwidth]{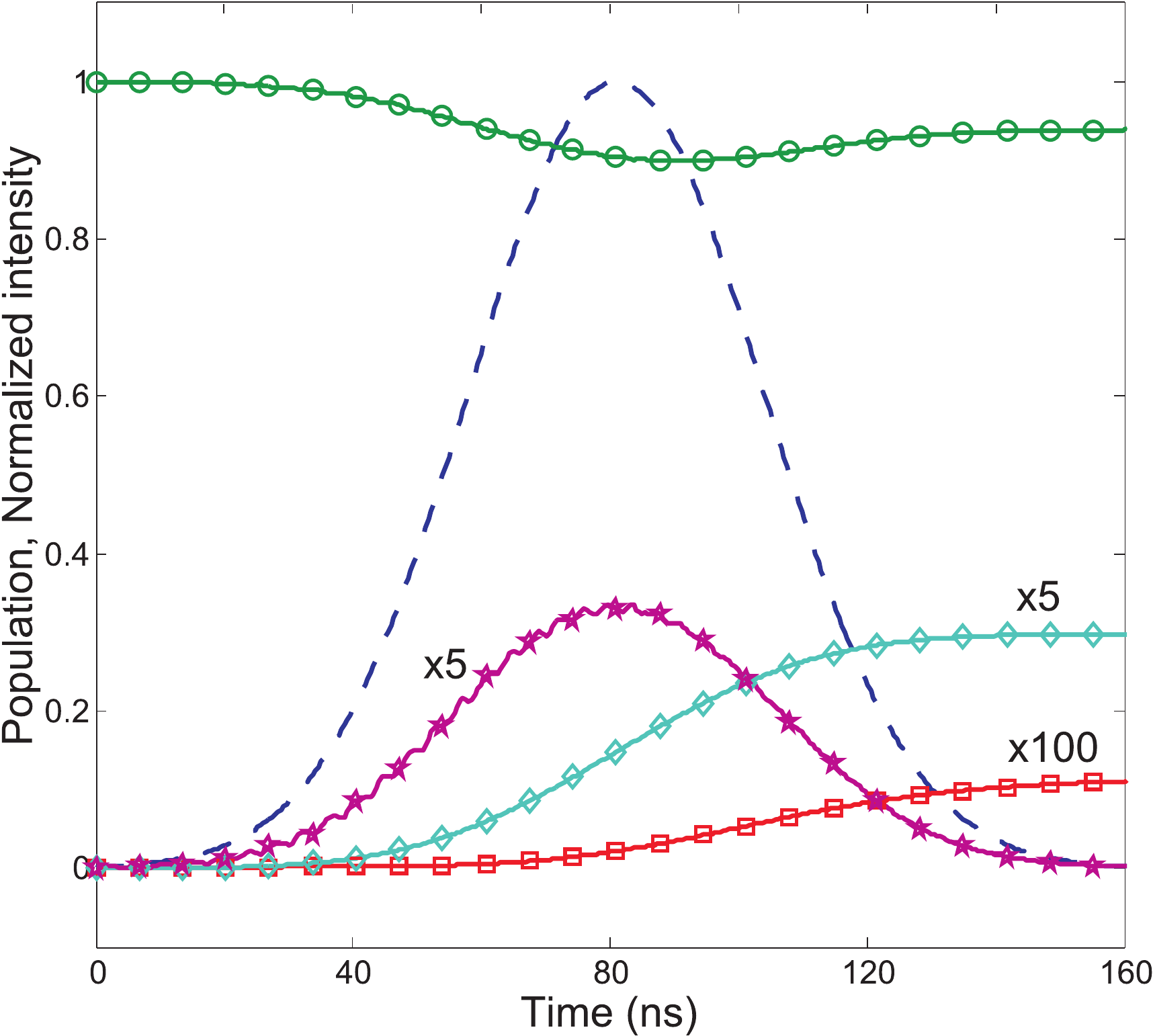}
    \end{minipage}\hfill
    \begin{minipage}[b]{0.38\linewidth}
      \caption{\textbf{Population dynamics under nonlinear probing.} Results of simulations for a Gaussian pulse with a FWHM of 54ns, 5.7 million photons, a peak intensity of 4W/cm$^2$ and $\Delta=$ 462MHz, with an initially-polarized density matrix.
      Blue dashed line, pulse intensity (normalized). Symbols show total population for groups of states: Green circles, $\ket{1,1}$ state; red squares, $\ket{1,-1}$ and $\ket{1,0}$ states (100 $\times$ magnified); purple stars, excited states (5 $\times$ magnified); turqouise diamonds, $F = 2$ states (5 $\times$ magnified).
}
\label{fig:pulse}
    \end{minipage}
  \end{center}
\end{figure}

We solve to first order in $N_A$ as follows. We identify a solution to the
zero-atom equation ${\cal
D} \cEVec =0$ as the input field $\EInVec$.
Specifically, we take
%
%
 $\EInVec = \InputStr
\PolV \pulseshape(\tau) \beamshape(\bx)$  where $\PolV$ is the
unit vector in the $V$ direction, \begin{align*}\tag{S7} T(\tau) = \pi^{-1/4}
\tau^{-1/2} \exp[-t^2/2 \tau^2] \end{align*} is a Gaussian pulse with
 FWHM $2 \tau \sqrt{\ln 2}$, and \begin{align*}\tag{S8}  M(\bx) =
\sqrt{\frac{2}{\pi w ^2(z)}}  \exp[-r^2/w^2(z)] \exp[i \psi(r,z)]
\end{align*} where $w^2(z) = w_0^2(1+z^2/z_R^2)$, $z_R = \pi
w_0^2/\lambda$, and $\psi$ is the wave-front phase.  This
describes a Gaussian beam with effective area $A_0 \equiv \int dx
dy \, |M(x,y,0)|^2/|M({\bf 0})|^2 = \pi w_0^2/2$.

We find numerically the
solution to Eq. (\ref{Eq:oBEcm}) with $\cEVec =\EInVec$ as the
lowest-order atomic response $\rho^{(1)}$.  A representative case is shown in
Supp. Fig. \ref{fig:pulse}.    The evolving atomic state generates the field
 \begin{align*}\tag{S9}  \label{Eq:overlap}\cEVec^{(\rm
1)}(\bx,\tau) = \frac{k^2}{\varepsilon_0} \int d^3 x'\,
G(\bx,\bx') \cPVec^{(1)}(\bx',\tau)\end{align*} where $G$ is the Green
function for the PWE.

The detected signal is
 $\myexpect{\hat{S}_y} \equiv {\left(\hbar \omega Z_0\right)^{-1}}
\int  d^2 x d\tau\, \EInPar^* {\cal E}^{(1)}_{H} + c.c.$,
 where the spatial integral is taken over the surface of the
detector and subscripts $H,V$ indicate polarization components.
It can be shown, e.g., using Green function techniques$^{26}$, 
 that
\begin{align*}\tag{S10}
\int  d^2 x d\tau\, \EInPar^* {\cal E}^{(1)}_{H}  =  \frac{k }{2 i  \varepsilon_0}
\int d^3 x d\tau\,  \EInPar^* \cP^{(1)}_H
\end{align*}
and thus \begin{align*}\tag{S11}  \myexpect{\hat{S}_y} =  {\left(\hbar \omega
Z_0\right)^{-1}}\frac{k }{2 i \varepsilon_0 }  \int d^3 x \, n\int
d\tau\, \EInPar^*  \OneAtomPol^{(1)}_H  + c.c. \end{align*} while
$\myexpect{\hat{S}_x} = {\left(\hbar \omega Z_0\right)^{-1}} \int  d^2 x
d\tau\, |\EInPar|^2 = {\left(\hbar \omega Z_0\right)^{-1}} |\InputStr|^2 $.

\newcommand{\Erfc}{{\rm erfc}}
\newcommand{\MyErf}{F}
\newcommand{\Pone}{P_1}
\newcommand{\Ptwo}{P_2}
\newcommand{\GF}{{\cal G}}

{Independently determined values for the model parameters
$\sigma_L,\sigma_T,w_0$ and $\tau$ are used, leaving only $N_A$ as
a free parameter, found by fitting to the data.  We note that
$N_A$ determines the vertical position of the curve in Figure 3,
and has no effect on the sensitivity scaling.  In this sense, the
model confirms the scaling behaviour with no adjustable
parameters.}

Simulations indicate that loss of polarization in $F=1$, and thus
rotation signal, is mostly due to spontaneous decay into the F=2
ground level, as seen in Supp. Fig.~\ref{fig:pulse}.

\subsection{Sensitivity in time- and number-limited
scenarios} When time is limiting, the relevant sensitivity is
$\delta F_z \tau^{1/2}$, where $\delta F_z = \myexpect{\OpF_z} \delta \phi/\phi$
as in Equation 1, and the measurement duration $\tau$ is
$\tau_{\rm L} = \SI{40}{\micro s}$ or $\tau_{\rm NL} =
\SI{54}{ns}$ for the linear or nonlinear measurement,
respectively.
The sensitivity can be calculated from the measured values
$A(\Delta_L)=3.3\times 10^{-8}$, and $B(\Delta_0)=3.8\times
10^{-16}$,
 using
$\delta\phi=N^{-1/2}/2$, $\phi_L/\myexpect{\OpF_z}= A(\Delta_L)/2$ and
$\phi_{NL}/\myexpect{\OpF_z}=B(\Delta_0) N_{\rm NL}/2$.  We find $\delta
F_z^{(\rm L)} \tau_{\rm L}^{1/2} = 1.9 \times 10^{5}
\PerRtHz N_{\rm L}^{-1/2}$, and $\delta F_z^{(\rm NL)} \tau_{\rm
NL}^{1/2} = 6.1 \times 10^{11}\PerRtHz
N_{\rm NL}^{-3/2}$.  Given an equal number of photons $N_{\rm L} =
N_{\rm NL}=N$, the nonlinear technique surpasses the linear at
$N=3.2\times 10^6$, well within the super-Heisenberg portion of
the curve in Figure 3. 
In contrast, when time is not a limited resource, the sensitivity-per-measurement is $\delta F_z^{(\rm L)} =   3 \times 10^7 N_{\rm L}^{-1/2} $, and $\delta F_z^{(\rm NL)}  = 2.6 \times 10^{15} N_{\rm NL}^{-3/2}$. Extrapolating, the nonlinear technique would
surpass the linear at $N_{\rm NL}=8.7\times 10^7$, which is
however outside the super-Heisenberg portion of the curve in
Figure 3.

\section*{Supplementary Notes}
\begin{enumerate}
\item[25.] Windpassinger, P. J. {\it{et al.}} Ultra low-noise differential ac-coupled photodetector for sensitive pulse detection application.  {\it{Meas. Sci. Technol.}} {\bf{20}}, 0055201 (2009)

\item[26.] Mitchell, M. W. Parametric down-conversion from a wave-equation approach: Geometry and absolute brightness. {\it{Phys. Rev. A}} {\bf{79}}, 043835 (2009)
\end{enumerate}


\end{document}